\documentclass[a4paper,11pt]{article}
\usepackage{pos}

\title{Revisiting the PeVatron candidate MGRO J1908+06 with an updated H.E.S.S. analysis}
 \ShortTitle{MGRO J1908+06 with H.E.S.S.}

\author*[a]{D.~Kostunin}
\author[b]{L.~Mohrmann}
\author[a]{E.~de~Ona~Wilhelmi}
\author[b]{V.~Joshi}
\author[c]{A.~Mitchell}
\author[a]{S.~Ohm}
\author[d]{B.~Kh\'elifi}
\author[d,e]{L.~Giunti}
\author[f]{A.~Sinha}

\affiliation[a]{DESY, D-15738 Zeuthen, Germany}
\affiliation[b]{Friedrich-Alexander-Universit\"at Erlangen-N\"urnberg, Erlangen Centre for Astroparticle Physics, Erwin-Rommel-Str. 1, D 91058 Erlangen, Germany}
\affiliation[c]{Max-Planck-Institut f\"ur Kernphysik, P.O. Box 103980, D-69029 Heidelberg, Germany}
\affiliation[d]{Universit\'{e} de Paris, CNRS, Astroparticule et Cosmologie, F-75013 Paris, France}
\affiliation[e]{Institute for Research on the Fundamental Laws of the Universe (IRFU), Commissariat à l'énergie atomique (CEA), Universit\'e Paris-Saclay, F-91191 Gif-sur-Yvette, France}
\affiliation[f]{Laboratoire Univers et Particules de Montpellier, CNRS\\
  Universit\'{e} de Montpellier, F-34090 Montpellier, France}

\forColl{H.E.S.S.} 

\emailAdd{contact.hess@hess-experiment.eu}

\abstract{
Detecting and studying galactic gamma-ray sources emitting very-high energy photons sheds light on the acceleration and propagation of cosmic rays presumably created in these sources. Currently, there are few sources emitting photons with energies exceeding 100 TeV. In this work we revisit the unidentified source MGRO J1908+06, initially detected by Milagro, using an updated H.E.S.S. dataset and analysis pipeline. The vicinity of the source contains a supernova remnant and pulsars as well as molecular clouds. This makes the identification of the primary source(s) of galactic cosmic rays as well as the nature of the gamma-ray emission challenging, especially in light of the recent HAWC and LHAASO detection of the high energy tail of its spectrum. Exploiting the better angular resolution as compared to particle detectors, we investigate the morphology of the source as well as its spectral properties.
}

\FullConference{37$^{\rm{th}}$ International Cosmic Ray Conference (ICRC 2021)\\
		July 12th -- 23rd, 2021\\
		Online -- Berlin, Germany}

\usepackage{xcolor}
\usepackage{xparse}

\ExplSyntaxOn

\keys_define:nn { miguel/label }
 {
  label   .tl_set:N = \l_miguel_label_tl,
  unknown .code:n   = \clist_put_right:Nx \l_miguel_label_clist
                       { \l_keys_key_tl = \exp_not:n { #1 } }
 }
\clist_new:N \l_miguel_label_clist
\box_new:N \l_miguel_label_box
\box_new:N \l_miguel_label_image_box

\NewDocumentCommand{\xincludegraphics}{O{}m}
 {
  \tl_clear:N \l_miguel_label_tl
  \clist_clear:N \l_miguel_label_clist
  \keys_set:nn { miguel/label } { #1 }
  \tl_if_empty:NTF \l_miguel_label_tl
   {
    \miguel_includegraphics:Vn \l_miguel_label_clist { #2 }
   }
   {
    \hbox_set:Nn \l_miguel_label_image_box
     {
      \miguel_includegraphics:Vn \l_miguel_label_clist { #2 }
     }
    \hbox_set:Nn \l_miguel_label_box
     {
      \skip_horizontal:n { 3pt }
      \fcolorbox{black}{white}{\footnotesize \tl_use:N \l_miguel_label_tl}
     }
    \leavevmode
    \box_use:N \l_miguel_label_image_box
    \skip_horizontal:n { -\box_wd:N \l_miguel_label_image_box }
    \hbox_overlap_right:n
     {
      \box_move_up:nn
       {
        \box_ht:N \l_miguel_label_image_box - 
        \box_ht:N \l_miguel_label_box - 3pt
       }
       { \box_use_drop:N \l_miguel_label_box }
     }
    \skip_horizontal:n { \box_wd:N \l_miguel_label_image_box }
   }
 }

\cs_new_protected:Nn \miguel_includegraphics:nn
 {
  \includegraphics[#1]{#2}
 }
\cs_generate_variant:Nn \miguel_includegraphics:nn { V }

\ExplSyntaxOff

\begin{document}
\maketitle

\section{Introduction}
Mapping the sky in very-high-energy gamma-rays sheds light on the acceleration and propagation of Galactic cosmic rays, i.e. probing acceleration mechanisms in the vicinity of stellar objects, particle escape, etc.
One of the most interesting questions is to search for PeVatrons --- accelerators of cosmic-ray nuclei up to 1 PeV (the cosmic-ray knee), studying their population and acceleration cutoff.
The Galactic Center was the first detected PeVatron~\cite{Abramowski:2016mir}, recently H.E.S.S. detected another candidate HESS\,J1702$-$420~\cite{Giunti:2021ovn}.

Modern high-altitude particle arrays focused on gamma-ray detection brought intriguing results recently: 
HAWC has published a list of galactic sources with emission above 56~TeV, where a number of them feature emission above 100~TeV~\cite{Abeysekara:2019gov},
LHAASO has published the first dozen of its ultra-high energy gamma-ray sources~\cite{lhaaso}.
Some of them have been previously detected by H.E.S.S., e.g. eHWC\,J1825$-$134 (LHAASO\,J1825$-$1326), which HESS resolved into two sources; HESS J1825-137 and HESS J1826-130, demonstrating its superior angular resolution~\cite{Abdalla:2018qgt, Abdalla:2020dmz}.

This works describes a new analysis of HESS~J1908+063 (hereafter J1908+063), which was initially discovered by Milagro~\cite{Abdo:2007ad} and later confirmed by H.E.S.S.~\cite{Aharonian:2009je}, HAWC (eHWC\,J1907+063) and LHAASO (LHAASO\,J1908+0621).
Although the source was first detected more than a decade ago, its nature is still unidentified.
However, recent investigations of this region with different instruments have brought new insights as to its nature.
Besides recent results from ground gamma-ray detectors mentioned above, several multi-wavelength studies (including with Fermi-LAT and radio observations) have been conducted~\cite{Duvidovich:2019ldj,Crestan:2021ucl,Li:2021wzt} since J1908+063 has last been studied with H.E.S.S. in context of its Galactic Plane Survey \cite{H.E.S.S.:2018zkf}.
This progress motivates us to revisit this source using additional H.E.S.S. data not analysed previously with an updated reconstruction pipeline.

J1908+063 is spatially coincident with a number of stellar objects, which can potentially produce very-high-energy gamma rays.
Two main candidates have been suggested as the sources of emission: PSR\,J1907$+$0602~\cite{Collaboration:2010en} and SNR\,G40.5$-$0.5~\cite{Stil:2006pz} assuming pulsar wind nebula and molecular clouds interaction scenarios, respectively.
Recently discovered PSR\,J1907$+$0631~\cite{Lyne:2016kam}, possibly associated with SNR\,G40.5$-$0.5, is also energetically viable for gamma-ray production.
There are two more objects located at the northern edge of the source, which can also produce gamma rays and contribute to the J1908+063 emission: PSR\,J1906$+$0722~\cite{Clark:2015dza} and SNR\,3C397~\cite{2016ApJ...817...74L}.
The summary of their properties is given in Tab.~\ref{tab:objects}.

\begin{table}[b]
\centering
\begin{tabular}{lllll}
Object & $d$ (kpc) & $t_\mathrm{age}$ (kyr) & pulsar $\dot{E}$ (erg/s) & SNR radio size \\
\hline
PSR\,J1907$+$0602~\cite{Collaboration:2010en} & $3.2\pm0.6$ & 19.5 & $2.8\times 10^{36}$ & --- \\
PSR\,J1906$+$0722~\cite{Clark:2015dza} & $1.91$ & 49.2 & $1.02\times 10^{36}$ & ---\\
PSR\,J1907$+$0631~\cite{Lyne:2016kam} & $7.9$ & 11.2 & $5\times10^{35}$ & ---\\
SNR\,G40.5$-$0.5~\cite{Stil:2006pz} & $5.5$--$8.5$ & $20$--$40$ & --- & $22'$\\
SNR\,3C397~\cite{2016ApJ...817...74L} & $8$--$9$ & $1.35$--$5.3$ & --- & $4.5'\times2.5' $ \\
\end{tabular}
\caption{Objects in the vicinity of J1908+063, which can potentially produce very-high-energy gamma rays. The methods of age estimation are dependent on the object class.}
\label{tab:objects}
\end{table}

\section{Observation and analysis}
The High Energy Stereoscopic System (H.E.S.S.) is an array of four 12-m and one 28-m Imaging Atmospheric Cherenkov Telescopes (IACTs) located in the Khomas Highland in Namibia at an altitude of 1835~m.
It is capable of detecting VHE gamma-rays from energies of a few tens of GeV to 100 TeV.
The data used in this work were acquired prior to the installation of the 28-m telescope.

\begin{figure}[b]
	\centering
	\xincludegraphics[height=.35\textwidth, label=PRELIMINARY]{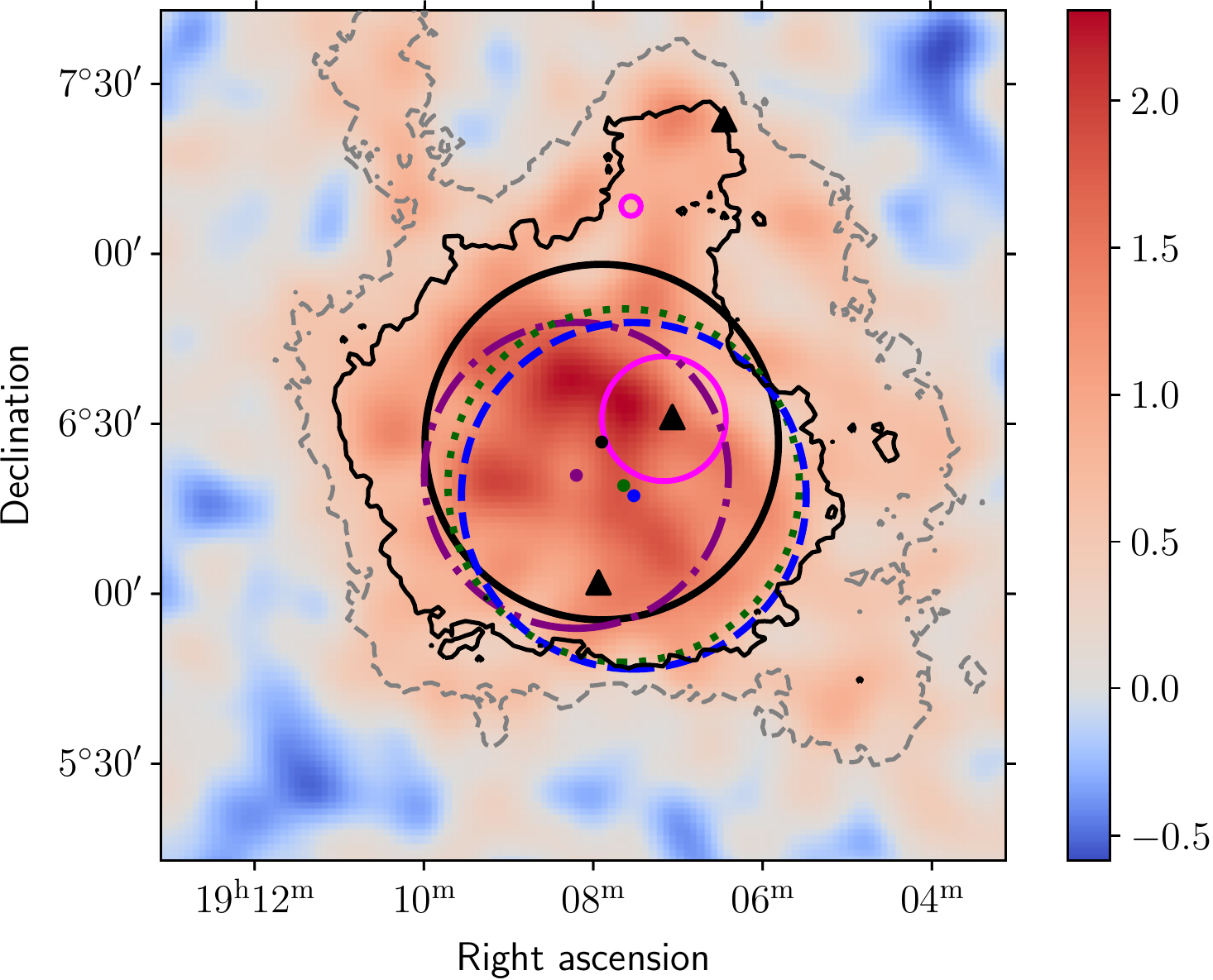}~~~~\xincludegraphics[height=.35\textwidth, label=PRELIMINARY]{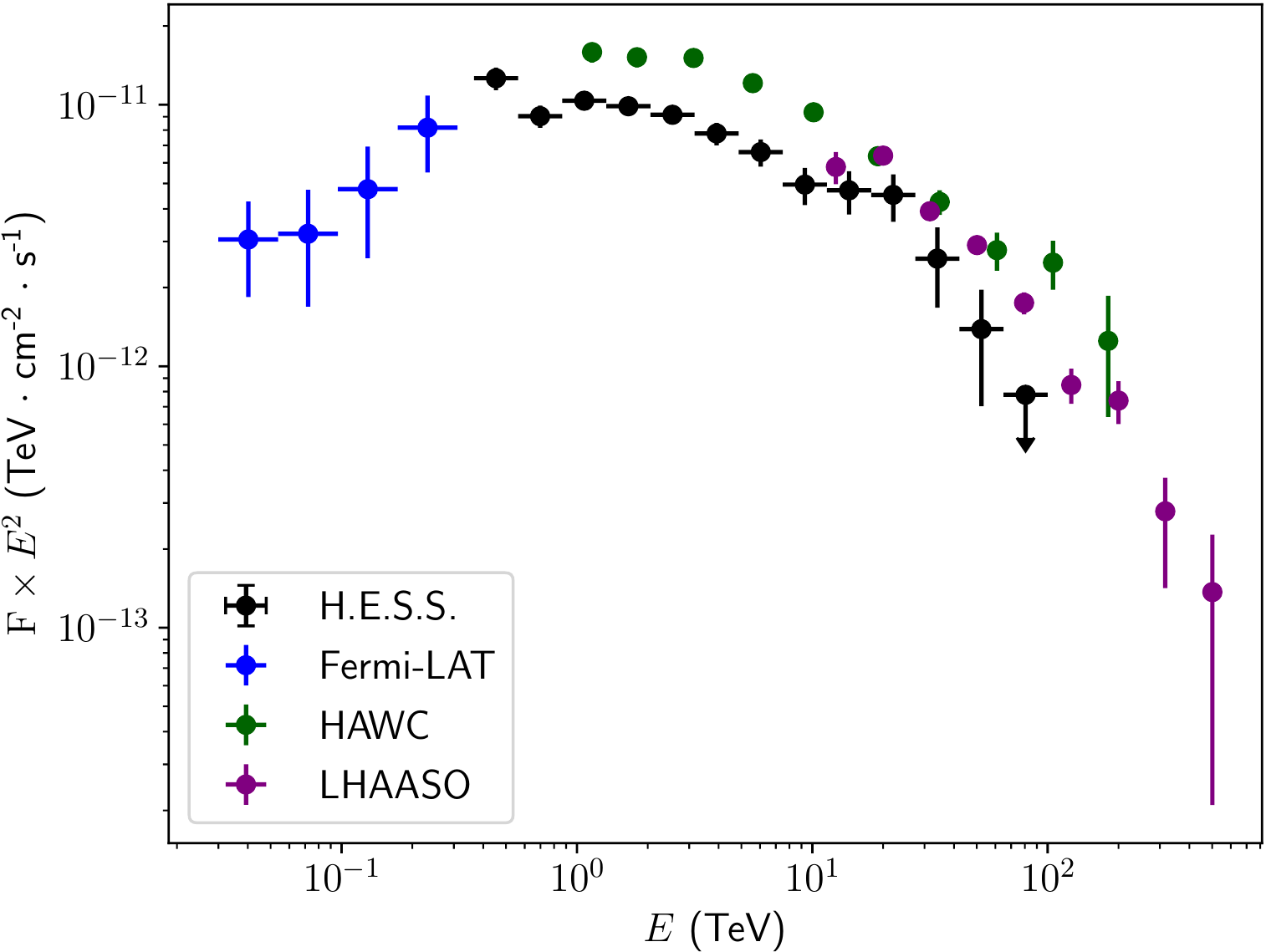}
	\caption{Comparison between single-component morphology and spectra reconstructed by different instruments with H.E.S.S. reconstruction obtained in this work.
	\textit{Left:}~H.E.S.S. excess map smoothed with radius of $0.07^\circ$. 
	Gray dashed and black solid contours correspond to $3\sigma$ and $5\sigma$ Li-Ma significance, respectively. 
	Circles correspond to morphology fits by H.E.S.S. 
	(black solid with $r=\sigma_\mathrm{source}$, 3D analysis with gammapy using Gaussian model), 
	Fermi-LAT (blue dashed with $r=r_\mathrm{disk}$, using disk model)~\cite{Li:2021wzt}, 
	HAWC (green dotted with $r=\sigma_\mathrm{source}$, using Gaussian model)~\cite{Abeysekara:2019gov} and LHAASO (purple dash-dotted radius corresponding to containment 68\% of source)~\cite{lhaaso}.
	Triangles indicate pulsars, magenta circles indicate supernova remnants (see Tab.~\ref{tab:objects}). 
	\textit{Right:} comparison of the spectra reconstructed by different instruments using spacial models described above.}
	\label{fig:comp}
\end{figure}

Compared to the latest H.E.S.S. publication including J1908+063~\cite{H.E.S.S.:2018zkf}, the present work uses additional data acquired since the last publication, which adds about 50\% more exposure (mostly in the northern part of the source).
The resulting livetime for the central part of the source is almost 80~hours after applying quality cuts.
For the low-level reconstruction we use events detected by four telescopes with maximum offset of $2^\circ$ from the center of the camera.
Direction and energy reconstruction and gamma/hadron separation is performed with methods described in Refs.~\cite{Parsons:2014voa,Ohm:2009nw}.

The high-energy analysis is performed with \texttt{gammapy} v0.17~\cite{gammapy:2017,gammapy:2019} with an energy threshold of $\sim0.365$~TeV and three-dimensional model of residual hadronic background.
For the cross-check with previous H.E.S.S. results and newest ones obtained by Fermi-LAT~\cite{Li:2021wzt}, HAWC~\cite{Abeysekara:2019gov} and LHAASO~\cite{lhaaso}, we fitted J1908+063 assuming single Gaussian component with spectrum described by power law.
The preliminary fitted values are (only statistical errors are included): ${\mathrm{R.A.}=286.975^\circ\pm0.024^\circ}$, ${\mathrm{dec.}=6.432^\circ\pm0.024^\circ}$, sigma of Gaussian $\sigma=0.524^\circ\pm0.018^\circ$.
The spectrum described by the parametrization $\phi(E) = \phi_0(E/\mathrm{TeV})^{-\Gamma}$ has the following values (only statistical errors are included): $\phi_0 = (1.02\pm0.05\,)\cdot10^{-11}$\,TeV\textsuperscript{$-$1}\,cm\textsuperscript{$-$2}\,s\textsuperscript{$-$1}, $\Gamma=2.294\pm0.027$.
The comparison of morphological and spectral fits between different instruments is given in Fig.~\ref{fig:comp}.

\pagebreak

\begin{figure}[h!]
	\xincludegraphics[width=.32\textwidth, label=PRELIMINARY]{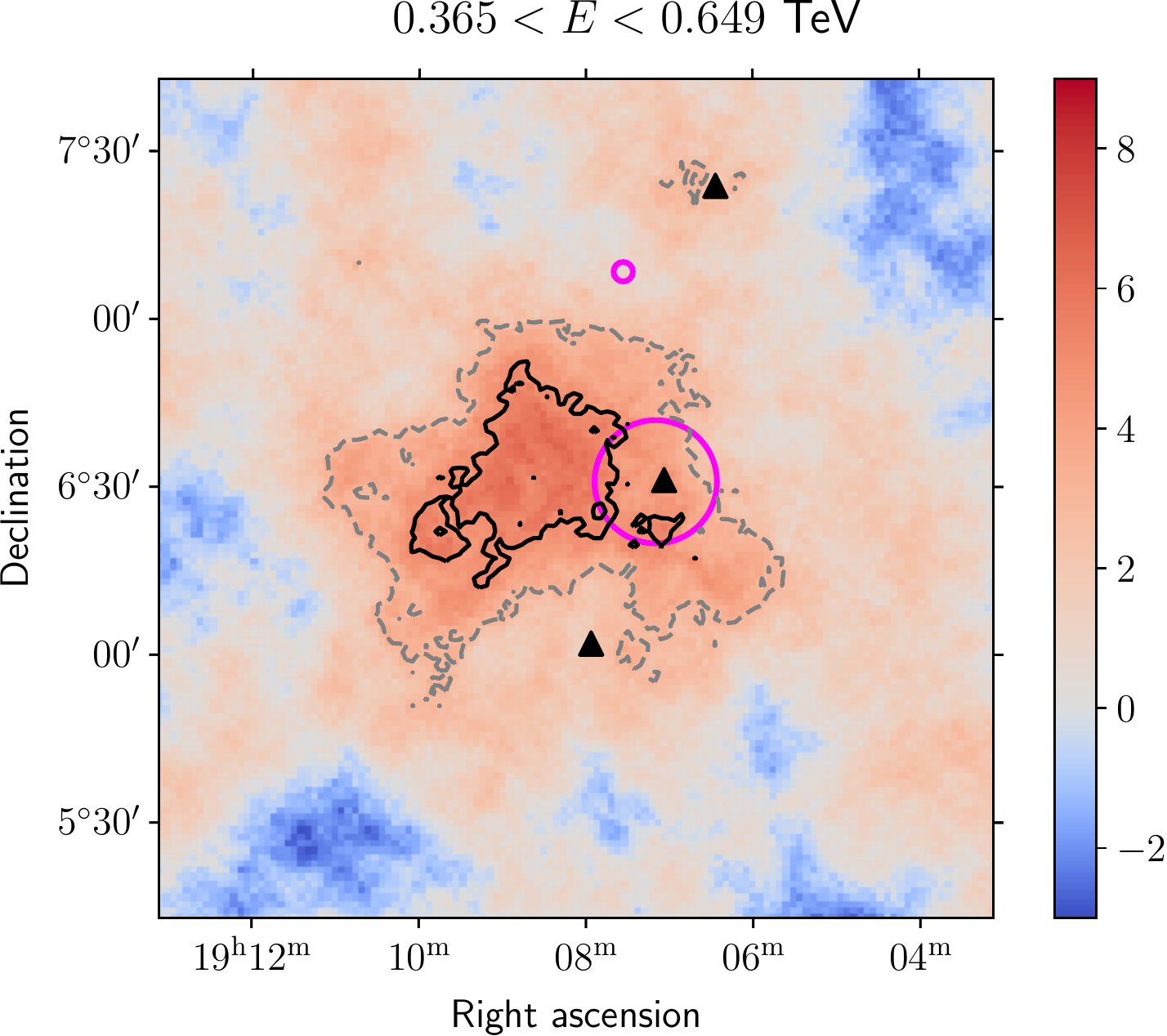}
	\xincludegraphics[width=.32\textwidth, label=PRELIMINARY]{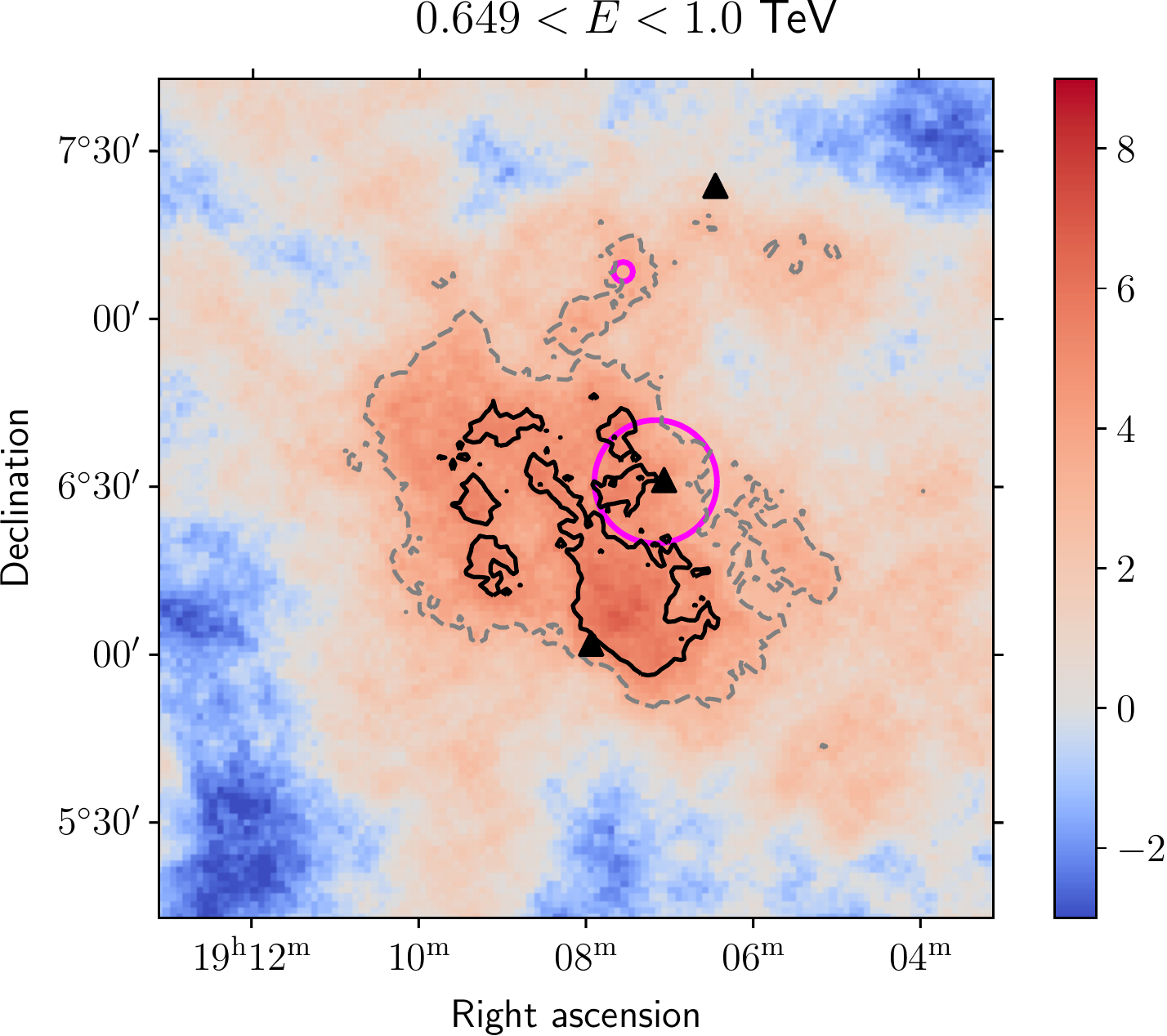}
	\xincludegraphics[width=.32\textwidth, label=PRELIMINARY]{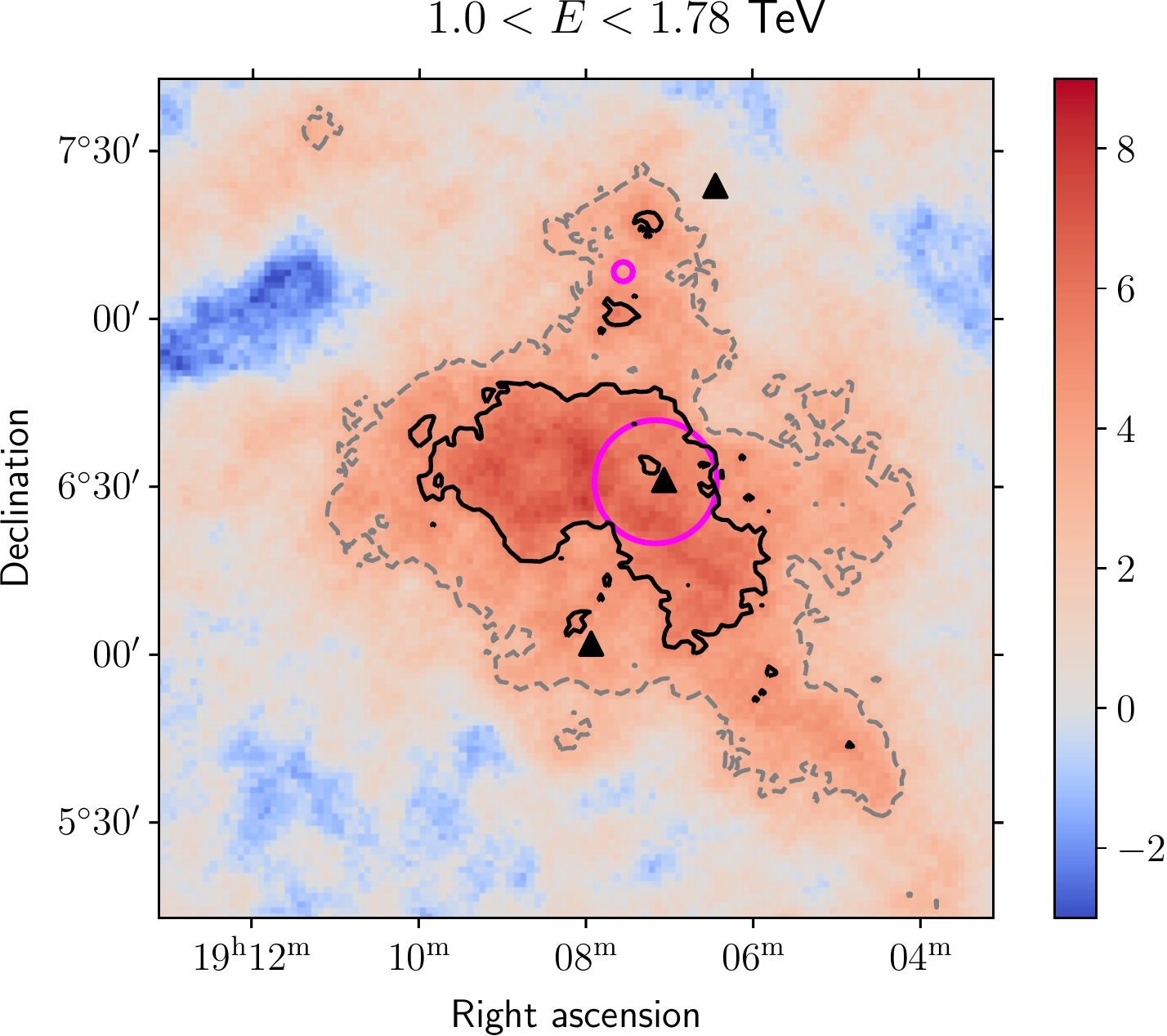}\\
	~\\
	\xincludegraphics[width=.32\textwidth, label=PRELIMINARY]{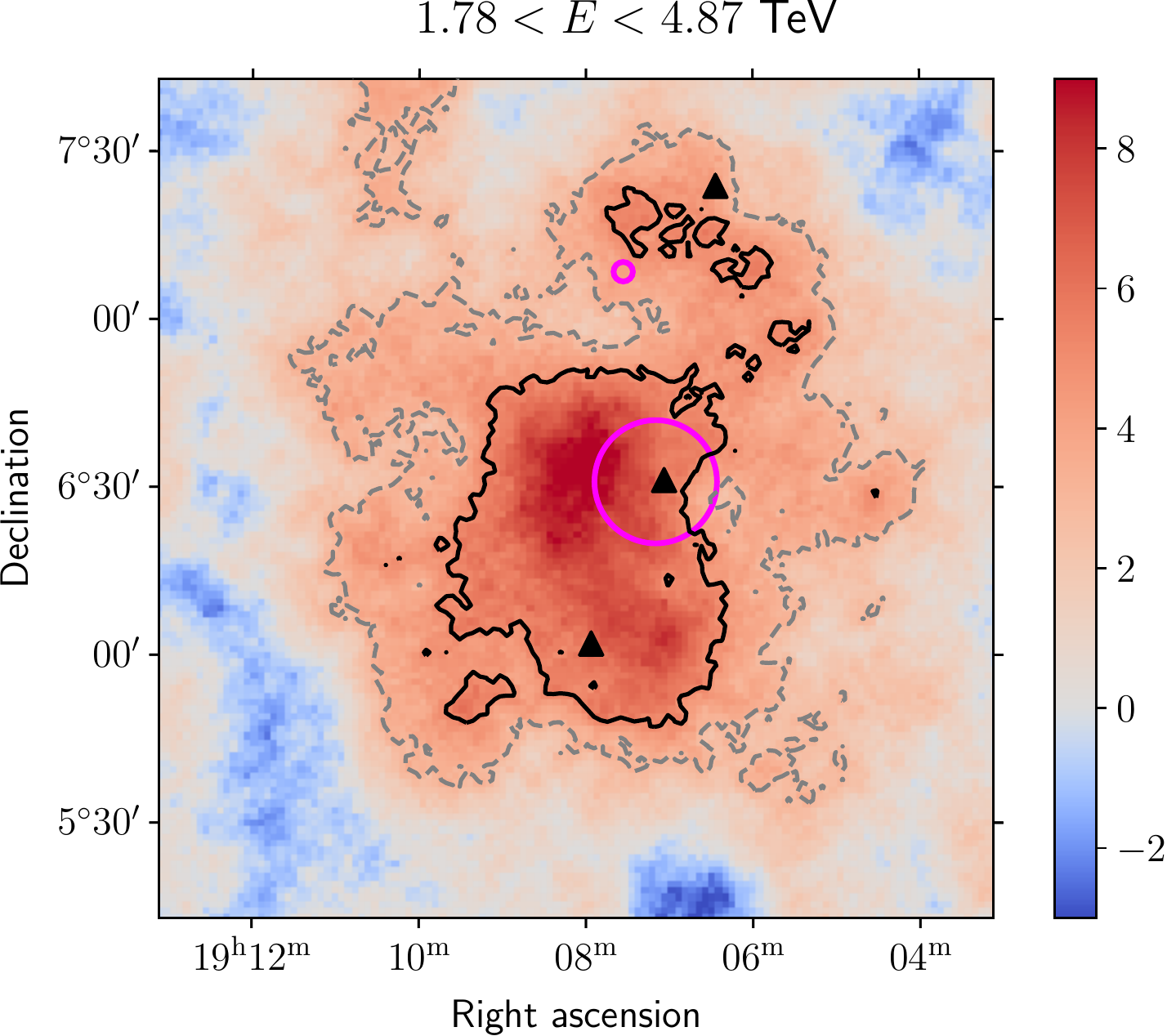}
	\xincludegraphics[width=.32\textwidth, label=PRELIMINARY]{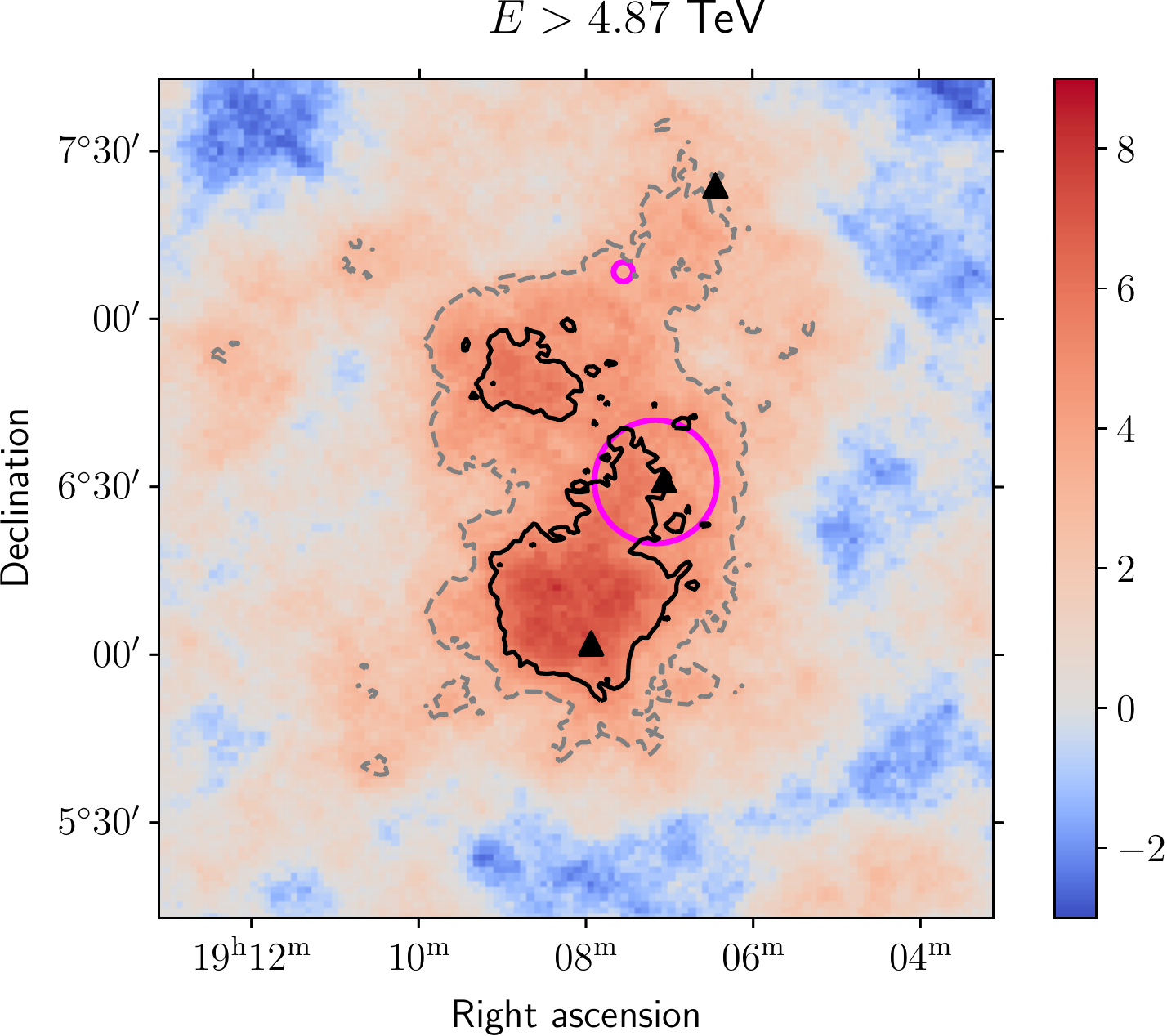}
	\xincludegraphics[width=.32\textwidth, label=PRELIMINARY]{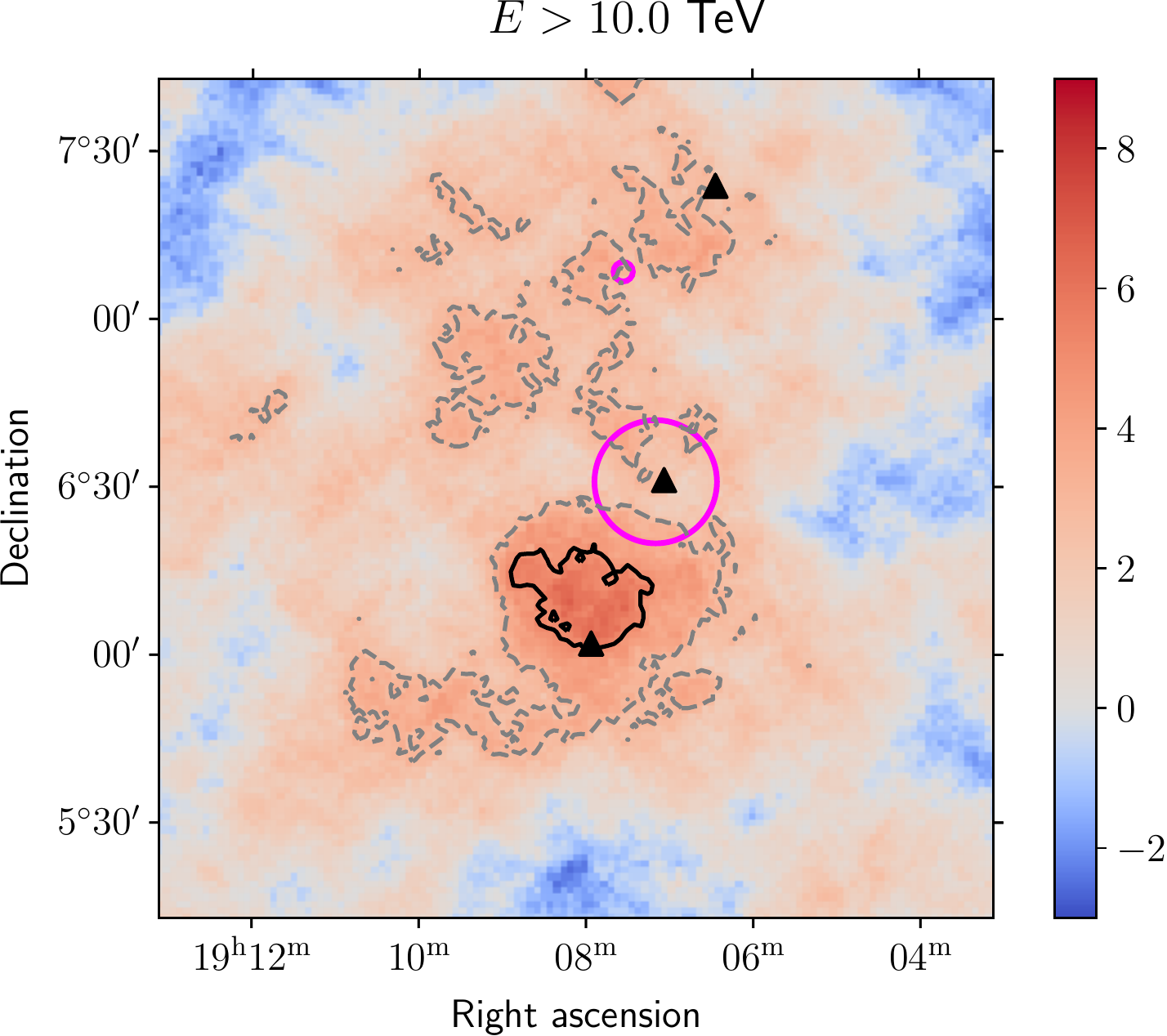}
	\caption{Significance maps reconstructed for 6 energy bands using Li-Ma definition~\cite{Li:1983fv}. Gray dashed and black solid contours correspond to 3 and 5 sigma significance, respectively. All color scales are defined with the bounds $(-3; 8)$ sigma in order to simplify comparison. See Fig.~\ref{fig:comp} for further details on the legend.}
	\label{fig:significance}
\end{figure}

The next step in our analysis is to probe for energy-dependent features of the source.
We have chosen six energy bins, four of which --- $(0.365,\,0.649)$, $(0.649,\,1.0)$, $(1.0, 1.78)$, $(1.78, 4.87)$~TeV --- contain roughly the same number of excess events (about 1100), and two high-energy regions: $E>4.87$~TeV (420 excess events) and $E>10$~TeV (190 excess events) still featuring significant detection of the source.
The significance maps are given in Fig.~\ref{fig:significance}.
We have searched for dependence of single-component fit parameters on energy in these bins, however they stay constant within uncertainties.
Taking into account the complex surroundings of J1908+063 we plan to perform morphology fits using different models and number of components.

\begin{table}[b]
\centering
\begin{tabular}{lllllll}
Region & R.A. & Dec. & Radius & Excess & $\phi_0$ ($10^{-12}\,\mathrm{TeV}^{-1}\mathrm{cm}^{-2} \mathrm{s}^{-1}$) & $\Gamma$ \\
\hline
A & $286.87^\circ$ & $6.12^\circ$ & $0.4^\circ$--$0.8^\circ$ & 1413.9 & $1.89\pm0.12$ & $2.43\pm0.06$ \\
B & $286.92^\circ$ & $6.17^\circ$ & $0.2^\circ$ & 498.3 & $0.637\pm0.071$ & $2.12\pm0.07$\\
C & $286.65^\circ$ & $7.20^\circ$ & $0.2^\circ$ & 197.6 & $0.265\pm0.061$ & $2.12\pm0.13$\\
\end{tabular}
\caption{Preliminary geometrical and spectral properties for the regions A, B, C reconstructed for the power-law model. Confidence regions include statistical uncertainties only.}
\label{tab:spectra}
\end{table}

\begin{figure}[h!]
	\centering
	\xincludegraphics[height=.35\textwidth, label=PRELIMINARY]{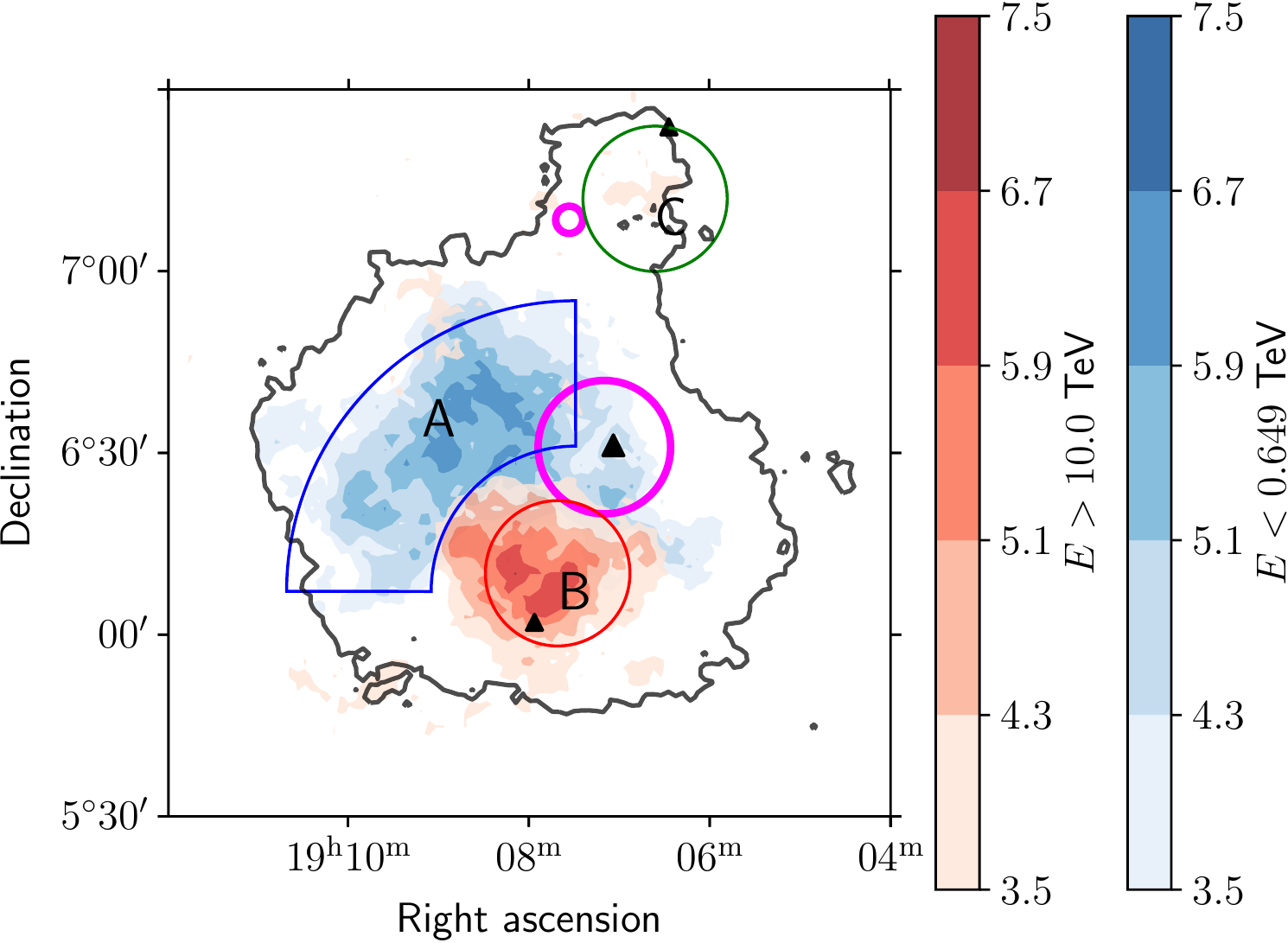}~~~~\xincludegraphics[height=.35\textwidth, label=PRELIMINARY]{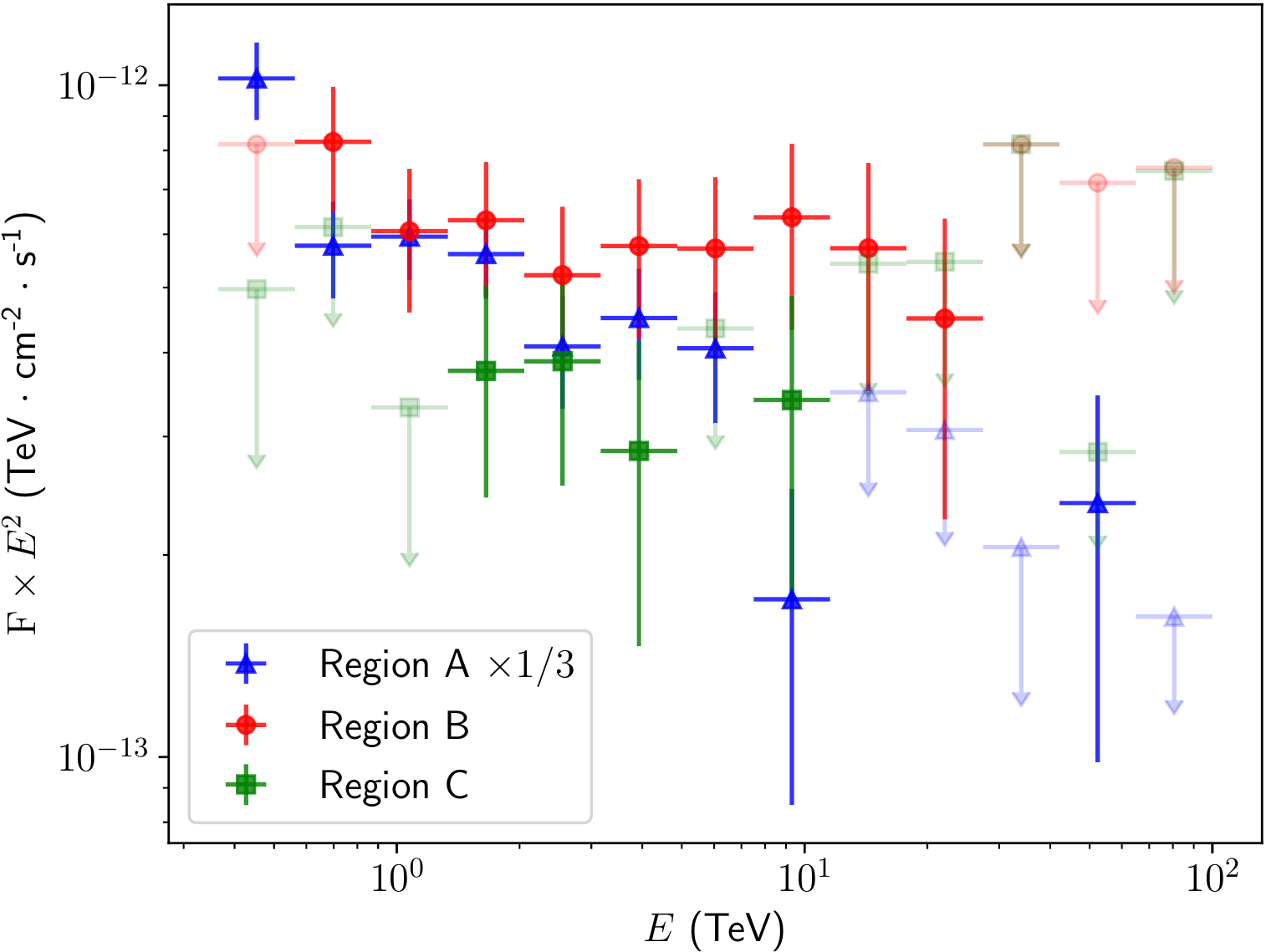}
	\caption{Comparison of spectra reconstructed from the different parts of the source. 
	Region A (blue segment) in the North-West part of the source, roughly equally distanced from pulsars and supernova remnant, 
	Region B (red circle) in the southern part of the source nearby PSR\,J1907+0602 and 
	Region C (green circle) in the northern part of the source nearby PSR\,J1906+0722. 
	\textit{Left:} configuration of the regions, black contour represents $5\sigma$ Li-Ma significance contour for all-energy emission, 
	blue and line contours are Li-Ma significance contours for low-energy ($E<0.649$\,TeV) and high-energy ($E>10$\,TeV) emission.
	\textit{Right:}~Spectra extracted from three regions normalized by solid angle. 
	Spectrum for Region A is normalized for the solid angle to ease the comparison. See Fig.~\ref{fig:comp} for further details on the legend.}
	\label{fig:regions}
\end{figure}

In the present analysis we pre-defined three regions based on possible origins of the gamma-ray emission with astrophysical associations.
Fig.~\ref{fig:regions} shows the geometry configuration of the regions and their spectra, while Tab.~\ref{tab:spectra} shows the fitted parameters of power-law spectral model.
There is indication of spectral difference across the region, but this remains to be confirmed with further tests.

\begin{figure}
\centering
\xincludegraphics[width=1.0\textwidth, label=PRELIMINARY]{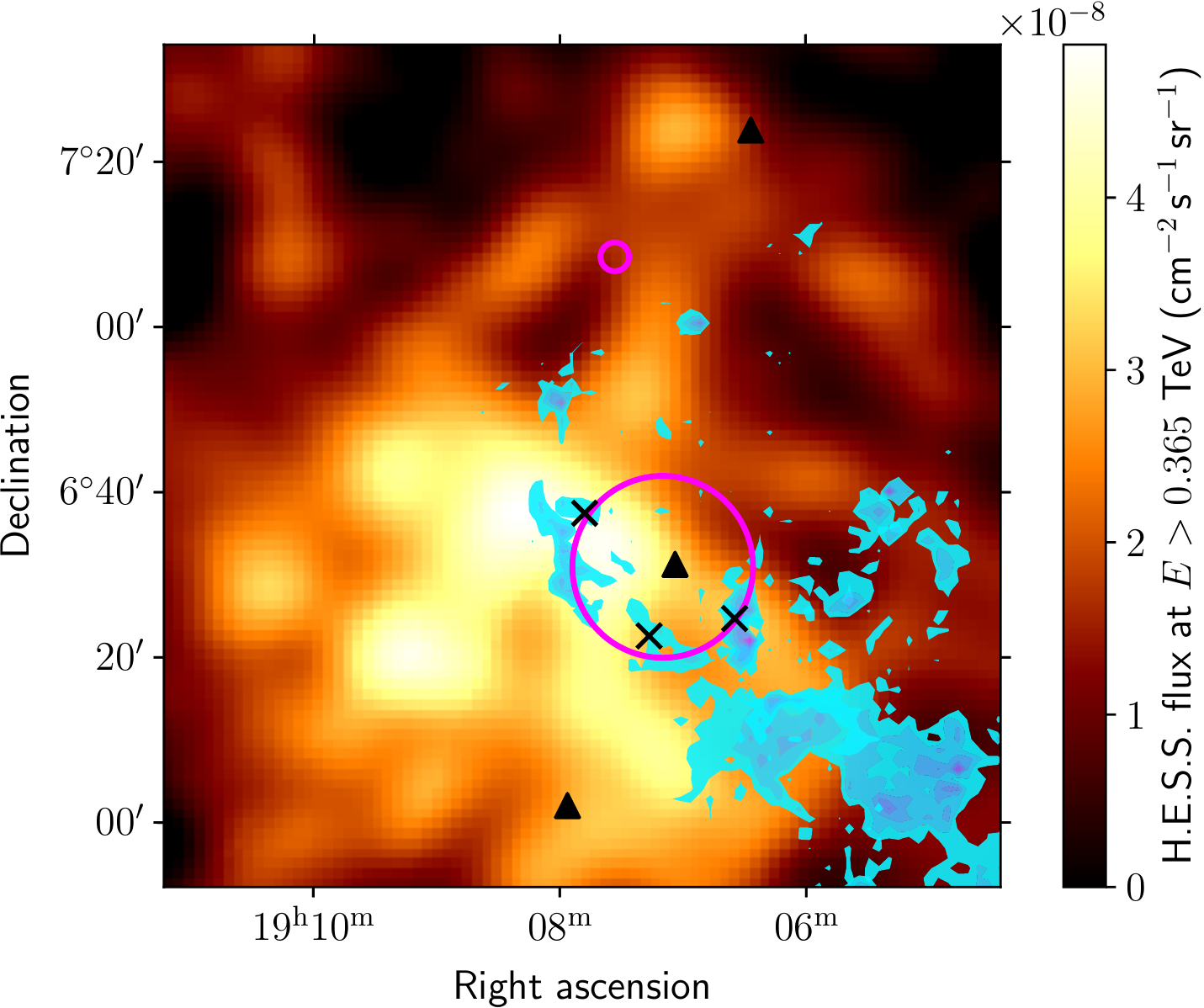}
\caption{H.E.S.S. flux maps combined with \textsuperscript{12}CO measurements and possible interaction with SNR\,G40.5$-$0.5 as suggested in Ref.~\cite{Duvidovich:2019ldj}.
Filled contours (cyan) show molecular clouds at different distances, crosses indicate possible hadronic interactions (see Ref.~\cite{Duvidovich:2019ldj}).}
\label{fig:mwl}
\end{figure}

As an intermediate result we produce the integral flux maps assuming a power law with a spectral index of $2.3$ and show them in combination with radio measurements from Ref.~\cite{Duvidovich:2019ldj}, see Fig.~\ref{fig:mwl}.
We see no striking spatial correlation between very-high-energy gamma-ray emission detected by H.E.S.S. and radio emission as detected in Ref.~\cite{Duvidovich:2019ldj}.

\section{Conclusion and discussion}
The vicinity of J1908+063 contains plenty of stellar objects as well as dense molecular clouds, which can generate very-high-energy gamma emission.
In this work we have revisited this source with an extended H.E.S.S. dataset and new analysis pipeline using \texttt{gammapy} for high-level analysis, which has given a significant detection in six energy bins and allowed for the testing of energy-dependent morphology.

The single-component fit obtained using 3D analysis in \texttt{gammapy} is in very good agreement with the previously published H.E.S.S. results.
It is worth noting that the spectral reconstruction by this new analysis technique does not require containment correction (as described in Ref.~\cite{H.E.S.S.:2018zkf}), since it takes into account the entire source and assumes pre-defined morphology (Gaussian shape in our case), which is important for studying of very extended sources.

This is an another cross-check and validation of  \texttt{gammapy}, a toolset for the future gamma-ray observatories, which has started to be used in H.E.S.S. recently~\cite{Mohrmann:2019hfq, Collaboration:2021fro, Giunti:2021ovn}.
~\\
~\\
~\\
~\\
~\\
~\\
The extended exposure for J1908+063 and surroundings sheds more light on the composition of this densely populated region.
We see hints on emission in the Northern part of the source.
This emission can be potentially associated with PSR\,J1906$+$0722 and SNR\,3C397.
To resolve possible contributions from these sources we will perform additional morphology fits, since all sources are in close proximity and featuring contamination from each other.
On the other hand these data are not sufficient to claim the origin of emission in the central part of the source, resolve more than one component or see energy-dependent morphology with high statistical significance.
The location of the brightest spots are in agreement within uncertainties and can be explained by inhomogeneous exposure and uncertainties in background estimation.
The same can be applied to the differences between spectral indices of Region A and Region B.

The results obtained point to the importance of investigating these kinds of sources with IACTs: while both Fermi-LAT and LHAASO brought interesting results on the lower and higher energy tails of the gamma spectra, only Cherenkov telescopes featuring higher angular resolution are able to distinguish between different contributions.
We will continue investigations of this source including multi-component fitting and modelling with \texttt{GAMERA}~\cite{Hahn:2015hhw} and \texttt{naima}~\cite{Zabalza:2015bsa} software.

\section*{Acknowledgements}
\begin{small}
This research made use of \texttt{gammapy},\footnote{\url{https://www.gammapy.org}} a community-developed core Python package for TeV gamma-ray astronomy \citep{gammapy:2017, gammapy:2019}. The maps of molecular clouds on Fig.~\ref{fig:mwl} are kindly provided by authors of Ref.~\cite{Duvidovich:2019ldj}.
The H.E.S.S. acknowledgements can be found in:
\url{https://www.mpi-hd.mpg.de/hfm/HESS/pages/publications/auxiliary/HESS-Acknowledgements-2021.html}
\end{small}

\bibliographystyle{JHEP}
\bibliography{hess-references}

\clearpage
\section*{Full Authors List: \Coll\ Collaboration}
\scriptsize
\noindent
H.~Abdalla$^{1}$, 
F.~Aharonian$^{2,3,4}$, 
F.~Ait~Benkhali$^{3}$, 
E.O.~Ang\"uner$^{5}$, 
C.~Arcaro$^{6}$, 
C.~Armand$^{7}$, 
T.~Armstrong$^{8}$, 
H.~Ashkar$^{9}$, 
M.~Backes$^{1,6}$, 
V.~Baghmanyan$^{10}$
% ...

\medskip

\noindent
$^{1}$University of Namibia, Department of Physics, Private Bag 13301, Windhoek 10005, Namibia\\
$^{2}$Dublin Institute for Advanced Studies, 31 Fitzwilliam Place, Dublin 2, Ireland\\
$^{3}$Max-Planck-Institut f\"ur Kernphysik, P.O. Box 103980, D 69029 Heidelberg, Germany\\
$^{4}$High Energy Astrophysics Laboratory, RAU,  123 Hovsep Emin St  Yerevan 0051, Armenia\\
$^{5}$Aix Marseille Universit\'e, CNRS/IN2P3, CPPM, Marseille, France\\
$^{6}$Centre for Space Research, North-West University, Potchefstroom 2520, South Africa\\
$^{7}$Laboratoire d'Annecy de Physique des Particules, Univ. Grenoble Alpes, Univ. Savoie Mont Blanc, CNRS, LAPP, 74000 Annecy, France\\
$^{8}$University of Oxford, Department of Physics, Denys Wilkinson Building, Keble Road, Oxford OX1 3RH, UK\\
$^{9}$IRFU, CEA, Universit\'e Paris-Saclay, F-91191 Gif-sur-Yvette, France\\
$^{10}$Instytut Fizyki J\c{a}drowej PAN, ul. Radzikowskiego 152, 31-342 Krak{\'o}w, Poland\\
% ...

%
%
%

\end{document}